\documentclass{aip-cp}
\usepackage[numbers]{natbib}
\usepackage{rotating}
\usepackage{graphicx}

\usepackage[rm,small]{caption}
\makeatletter
\renewcommand\@biblabel[1]{#1.}
\renewenvironment{thebibliography}[1]
     {\section*{\refname}%
      \@mkboth{\MakeUppercase\refname}{\MakeUppercase\refname}%
      \list{\@biblabel{\@arabic\c@enumiv}}%
           {
            \settowidth\labelwidth{\@biblabel{#1}}%
            \labelwidth0.75cm
            \setlength{\labelsep}{\dimexpr 0.75cm - \labelwidth}
            \setlength{\leftmargin}{\dimexpr \labelwidth +\labelsep}
            \itemindent0.0cm
	    \advance\leftmargin\labelsep
            \@openbib@code
            \usecounter{enumiv}%
            \let\p@enumiv\@empty
            \renewcommand\theenumiv{\@arabic\c@enumiv}}%
      \sloppy
      \clubpenalty4000
      \@clubpenalty \clubpenalty
      \widowpenalty4000%
      \sfcode`\.\@m}
     {\def\@noitemerr
       {\@latex@warning{Empty `thebibliography' environment}}%
      \endlist}
\makeatother

\usepackage{epstopdf}

\begin{document}

\title{Concept of a Staged FEL Enabled by Fast Synchrotron Radiation Cooling of Laser-Plasma Accelerated Beam by Solenoidal Magnetic Fields in Plasma Bubble}

\author[aff1]{Andrei Seryi\corref{cor1}}
\author[aff2]{Zsolt Lesz}
\author[aff2,aff3,aff4]{Alexander Andreev}
\author[aff1]{Ivan Konoplev}

\affil[aff1]{John Adams Institute for Accelerator Science, and Physics Department, University of Oxford, Keble Road, Oxford OX1 3RH, UK.}
\affil[aff2]{ELI-ALPS, Dugonics Square 13, 6720 Szeged, Hungary.}
\affil[aff3]{Max-Born Institute, Berlin, Germany.}
\affil[aff4]{Sankt Petersburg State University, St. Petersburg, Russia.}
\corresp[cor1]{Corresponding author: andrei.seryi@adams-institute.ac.uk}

\maketitle

\begin{abstract}
A novel method for generating GigaGauss solenoidal fields in a laser-plasma bubble, using screw-shaped laser pulses, has been recently presented. Such magnetic fields enable fast synchrotron radiation cooling of the beam emittance of laser-plasma accelerated leptons. This recent finding opens a novel approach for design of laser-plasma FELs or colliders, where the acceleration stages are interleaved with laser-plasma emittance cooling stages. In this concept paper, we present an outline of what a staged plasma-acceleration FEL could look like, and discuss further studies needed to investigate the feasibility of the concept in detail.  
\end{abstract}

\section{INTRODUCTION}
As shown recently in \cite{our-paper}, GigaGauss and beyond solenoidal fields can be generated in the laser-plasma bubble, using screw-shaped high intensity laser pulses. The large fields are created by the laser pulse inducing the rotational motion of electrons around the shell of the plasma bubble. It has been noted \cite{our-paper} that this solenoidal field, semi-stationary in the reference frame of the laser pulse, can be used, in particular, for guiding and providing synchrotron radiation beam-emittance cooling for laser-plasma-accelerated electron and positron beams. This opens up novel opportunities for designing of the light sources, free electron lasers, and high energy colliders based on laser plasma acceleration. 

Building up on the work and ideas for the new approach for FEL and collider design presented in \cite{our-paper}, we discuss a novel concept of a possible staged FEL, based on plasma acceleration, exploiting the properties of plasma bubble enabled by screw-shaped laser pulse, and relying on the TRIZ methodology \cite{Altshuller-book,Seryi-book} to create a detailed concept.

\section{CONCEPT OF A STAGED FEL}

\begin{figure}[h]
  \centerline{\includegraphics[width=\textwidth]{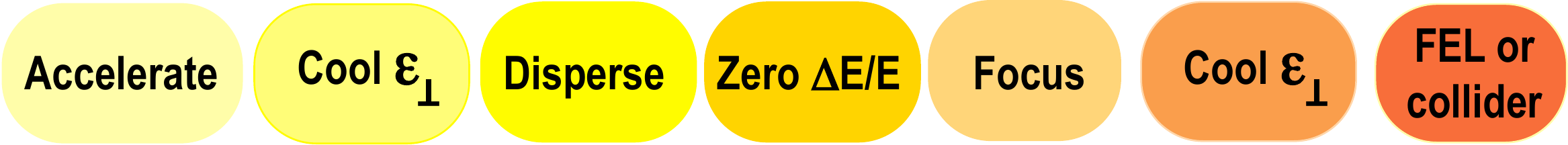}}
  \caption{Stages of Laser-Plasma Staged FEL: acceleration, cooling transverse degrees of freedom, dispersing the beam, monochromatization, focusing, one more transverse cooling stage, and sending the cooled beam to an FEL undulator or, for use in a collider.}
\label{all-scheme}
\end{figure}

The concept of a staged laser-plasma FEL is shown in Figure \ref{all-scheme}.

As discussed in \cite{our-paper}, a screw-shaped laser pulse creates longitudinal magnetic field in the plasma bubble and this can result in SR and cooling of the beam emittance. This by itself, however, is not sufficient to create a beam suitable for lasing in an FEL, as plasma-accelerated-beams typically have a large (several percent) energy spread. Correspondingly, the staged FEL shown in Fig.\ref{all-scheme} also includes the stages for monochromatization of the beam. Let's now look at the action of each stage in detail. 

\begin{figure}[h]
  \centerline{\includegraphics[width=0.39\textwidth]{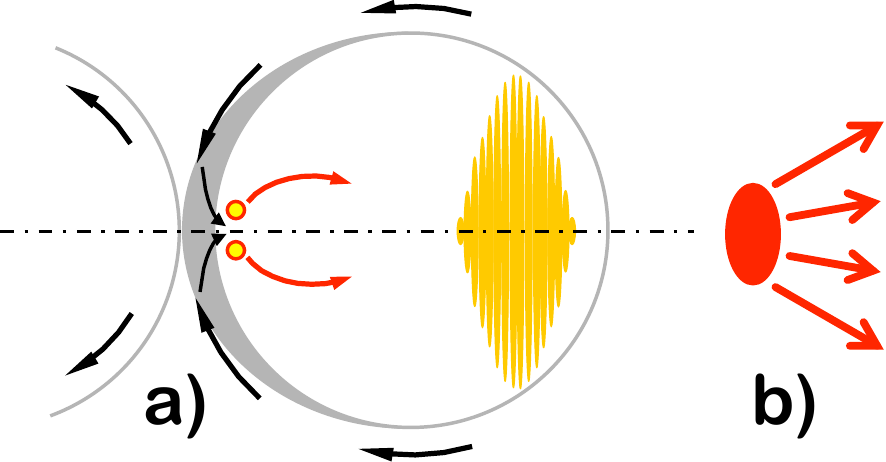}}
  \caption{Acceleration stage. Plasma bubble -- (a) shown in this case with a self-injected electron beam, and a schematic representation of the beam at the exit of the stage -- (b), which has a small size, but a relatively large divergence.}
\label{1-section-acceler}
\end{figure}
\begin{figure}[h]
  \centerline{
\includegraphics[width=0.27\textwidth]{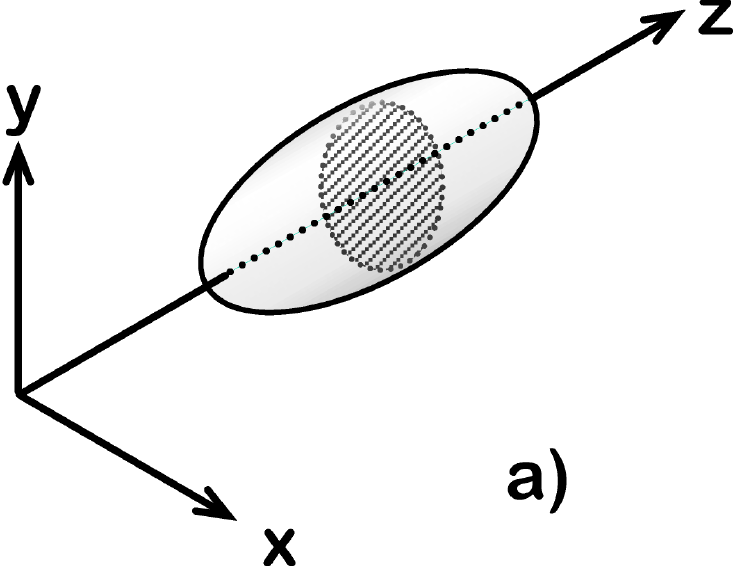}
\hspace{1pc}
\includegraphics[width=0.24\textwidth]{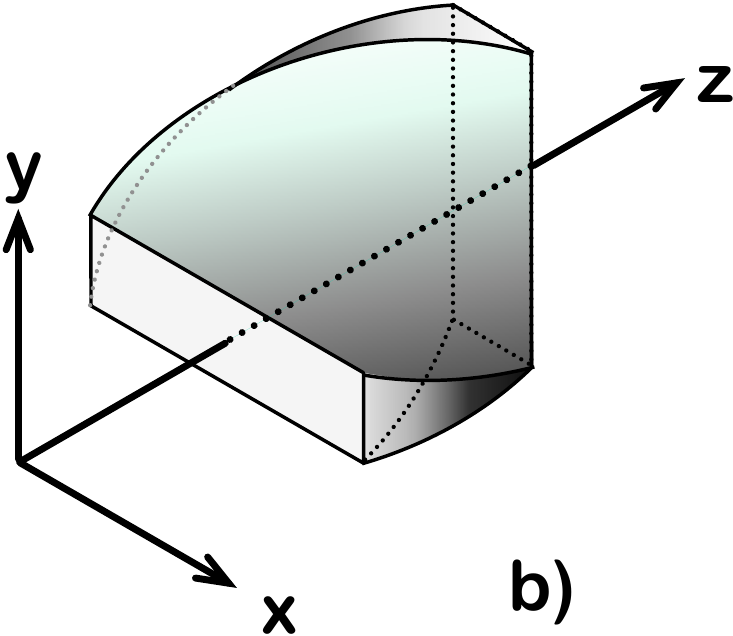}
\hspace{1pc}
\includegraphics[width=0.27\textwidth]{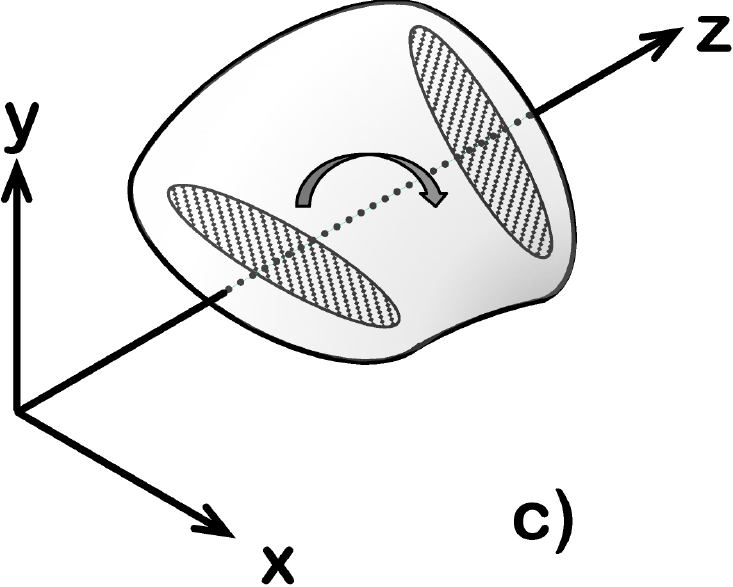}
}
  \caption{Laser pulse shapes used in the concept: standard round symmetrical laser pulse -- (a), screw-shaped laser pulse -- (b), slightly screw-shaped and asymmetrical laser pulse -- (c).} 
\label{round-pulse}
\end{figure}

The acceleration stage shown in Fig.~\ref{1-section-acceler} is the most standard. This is what has been studied in simulations as well as experimentally \cite{lwfa-1,lwfa-2} in recent years. A standard symmetrical laser pulse (see Fig.~\ref{round-pulse}(a)) is used to excite a plasma and create a plasma bubble. The beam at the section exit would have typically a GeV energy, small size, relatively large angular divergence, as well as few percent energy spread. 

The emittance of the accelerated GeV beam would need to be cooled, and, correspondingly, we use a screw-shaped laser pulse as shown in Fig.~\ref{round-pulse}(b) for the next section. 

\begin{figure}[h]
  \centerline{\includegraphics[width=\textwidth]{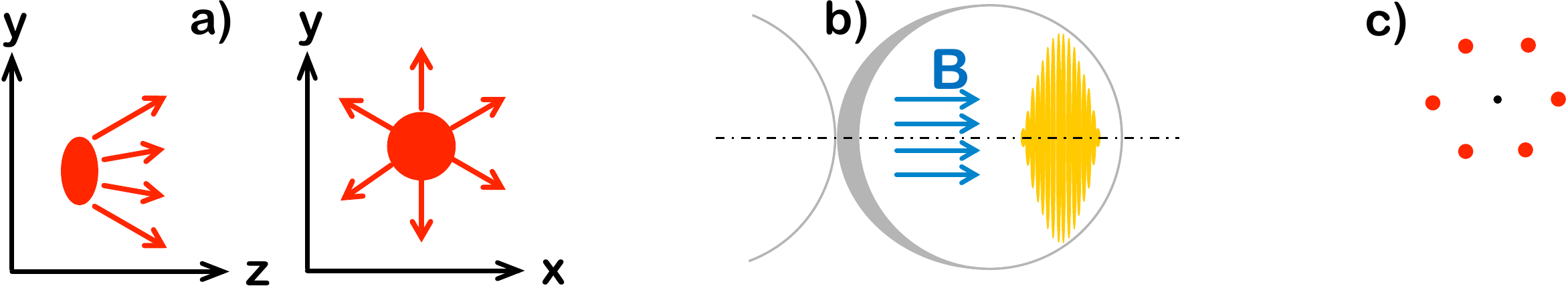}}
  \caption{Cooling Transverse stage. Schematics of the initial beam entering the stage -- (a), plasma bubble with large solenoidal field inside -- (b), schematics of the beam at the exit of the stage -- (c).}
\label{2-sect-cool-tr}
\end{figure}

\begin{figure}[h]
  \centerline{\includegraphics[width=0.5\textwidth]{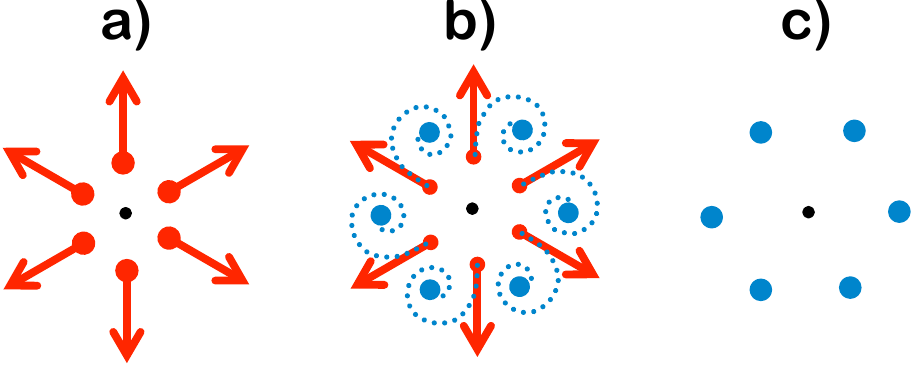}}
  \caption{Schematics of the beam before the cooling stage -- (a), inside of the bubble -- (b),  and  after the exit from the stage -- (c).}
\label{before-after}
\end{figure}

The first transverse cooling stage is shown in Fig.~\ref{2-sect-cool-tr}. 
The beam, which has a large angular divergence before the section, will reduce its angles, but, in general, will increase in diameter, as the electrons will spiral around individual centers, emitting synchrotron radiation and losing their energy (note that either this stage, or one of the following stages, may need to include re-acceleration, for for SR cooling to occur), but not reducing the size of the beam, as shown in Fig.~\ref{before-after}.  


Therefore, as a result of this cooling stage, a parallel beam with low angular spread but increased transverse size and the same energy spread will be produced. Correspondingly, we need to arrange a stage that will reduce the energy spread of the beam. This will be done via first dispersing the beam transversely, and then sending it to a special monochromatization stage.  

The stage that will disperse the beam in energies is using a standard laser pulse (Fig.~\ref{round-pulse}(a)), and the beam is injected into the bubble off center. The beam will feel transverse focusing forces in the bubble, and will receive a transverse kick that will depend on the energy of the electrons. One needs to note that a plasma lens could be used for the same purpose of dispersing or focusing the beam. 

\begin{figure}[h]
  \centerline{\includegraphics[width=\textwidth]{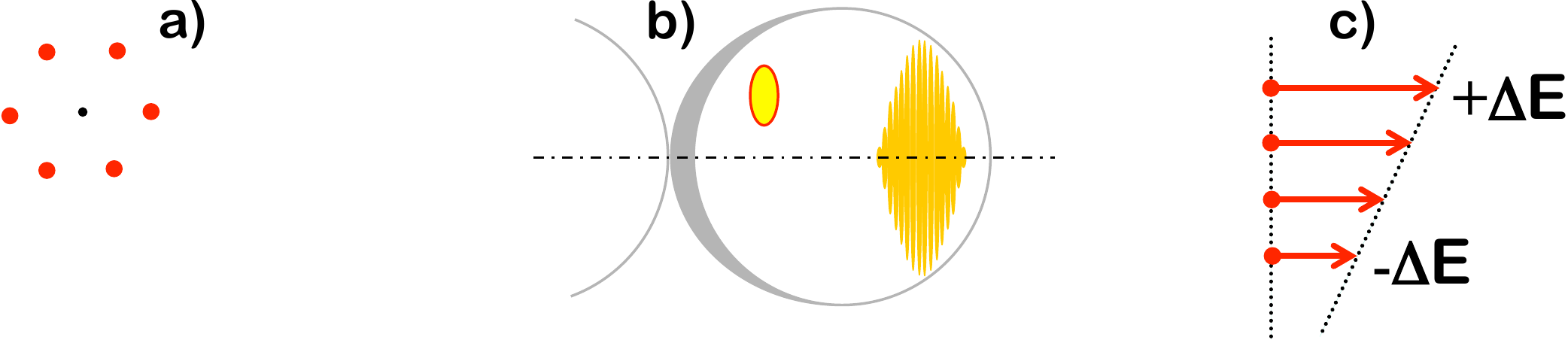}}
  \caption{Stage that disperses the beam transversely in energies. Schematics of the initial beam -- (a), plasma bubble where the beam is injected off-center -- (b), resulting beam at the exit of the section -- (c).}
\label{3-sect-disperse}
\end{figure}

\begin{figure}[h]
  \centerline{\includegraphics[width=0.8\textwidth]{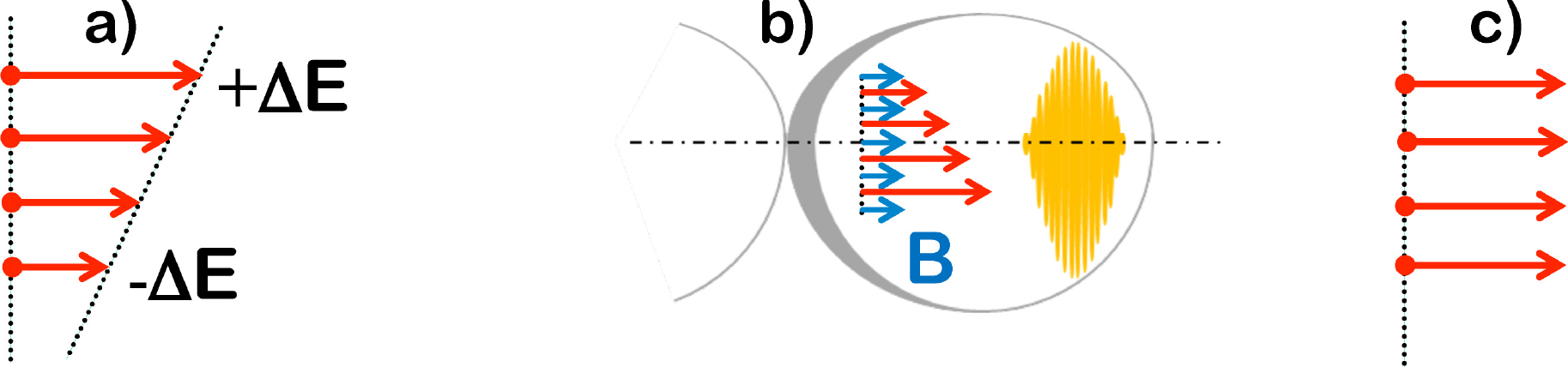}}
  \caption{Monochromatization stage. Schematics of the beam entering the stage -- (a), plasma bubble with some solenoidal field inside as well as transverse gradient of accelerating field -- (b), schematics of the beam at the exit of the stage -- (c).}
\label{4-sect-monochr}
\end{figure}

Correspondingly, at the exit of the dispersion section, we will have a parallel beam (low angular spread), but with similar size and energy spread, and, in addition, dispersed in energy. This beam will now enter the monochromatization stage shown in Fig.~\ref{4-sect-monochr}. 

For the monochromatization stage, we rely on a hypothesis that a slightly screw-shaped and also asymmetrical laser pulse (as shown in Fig.~\ref{round-pulse}(c)) will create a bubble that would have not only some solenoidal field, but also a transverse gradient of the accelerating field. The latter will provide compensation of the energy spread of the dispersed beam. Also, some solenoidal field is needed in the bubble, in order to keep the electrons circling around magnetic field lines, and not performing transverse betatron oscillations in the bubble, that would otherwise smear out the  monochromatization effect. In the process of optimization of the parameters of the monochromatization stage, we would need to take into account the possible {\bf{E}}$\times${\bf{B}} drift (where {\bf{E}} corresponds to the transverse focusing in the bubble) as well as possible {\bf{B}}$\times\nabla${\bf{B}} drift. A parameter range would need to be found where monochromatization works despite of these possible drifts or other effects. 

At the monochromatization section exit, correspondingly, we will have an accelerated low-energy-spread beam, which is also parallel (low angular spread), but still has large size. Reducing its size is done in the next two stages -- focusing and the second transverse cooling stage. 

\begin{figure}[h]
  \centerline{\includegraphics[width=0.85\textwidth]{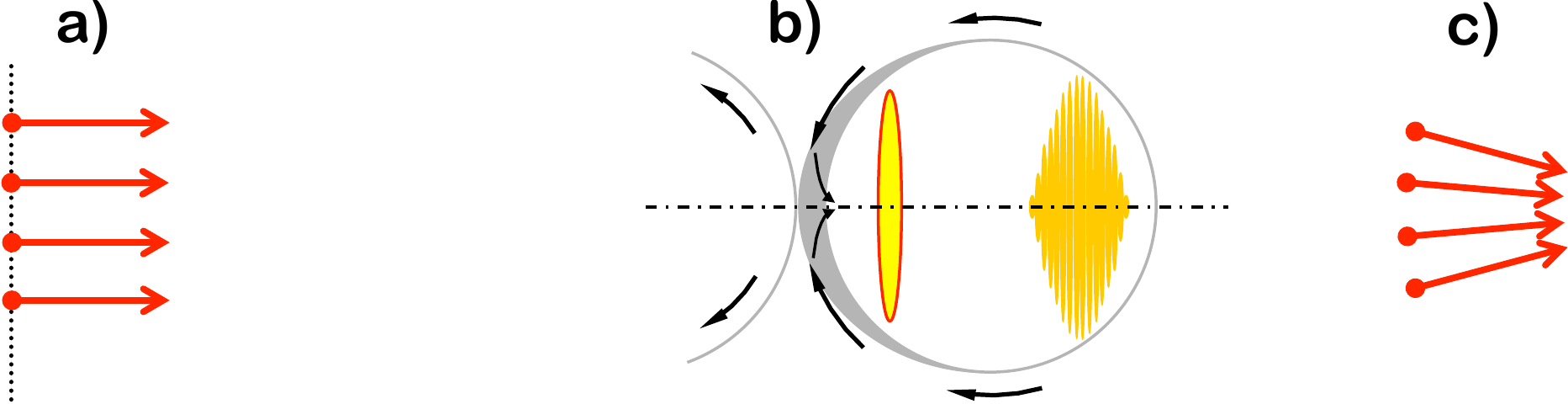}}
  \caption{Focusing stage. Schematics of the beam before the stage -- (a), plasma bubble -- (b), beam after the stage -- (c).}
\label{5-sect-focus}
\end{figure}

The focusing stage (see Fig.~\ref{5-sect-focus}) is driven by a standard symmetrical laser pulse, and it converts our beam into a converging beam that still has low energy spread and low angular spread (apart from the correlated convergence angles). One needs to note again that this focusing function can also be performed by a plasma lens. 

\begin{figure}[h]
  \centerline{\includegraphics[width=0.85\textwidth]{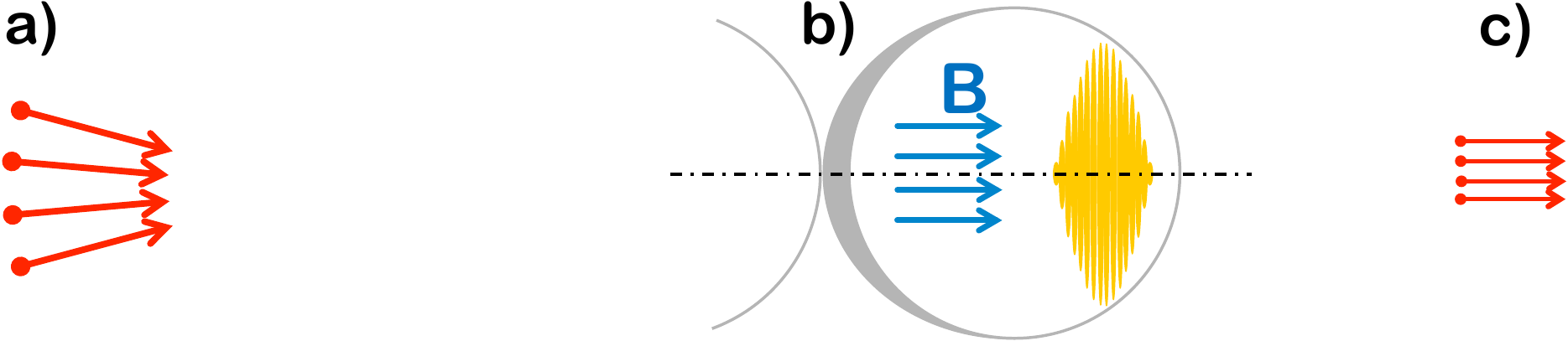}}
  \caption{Second transverse cooling stage. Schematics of the beam before the stage -- (a), plasma bubble -- (b), beam after the stage -- (c).}
\label{final-cool-tr}
\end{figure}

Following the focusing stage, one more transverse cooling stage is inserted, as shown in Fig.~\ref{final-cool-tr}, in order to cool down the angles corresponding to converging trajectories of the electrons of the beam. 

\begin{figure}[h]
  \centerline{\includegraphics[width=0.95\textwidth]{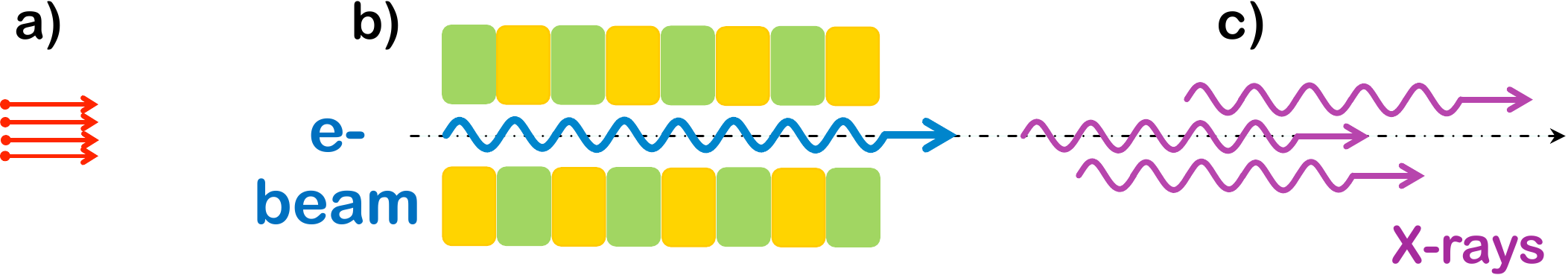}}
  \caption{FEL stage. The beam before the stage -- (a), undulator of an FEL -- (b), and generated radiation -- (c).}
\label{stage-fel}
\end{figure}

As a result, the beam will have small energy spread, small size and low angular spread, and will be suitable for sending to an undulator of the FEL stage for lasing, as shown in Fig.~\ref{stage-fel}, or to be sent to the same sequence again, for further acceleration and cooling, for collider applications.

\begin{figure}[h]
  \centerline{\includegraphics[width=\textwidth]{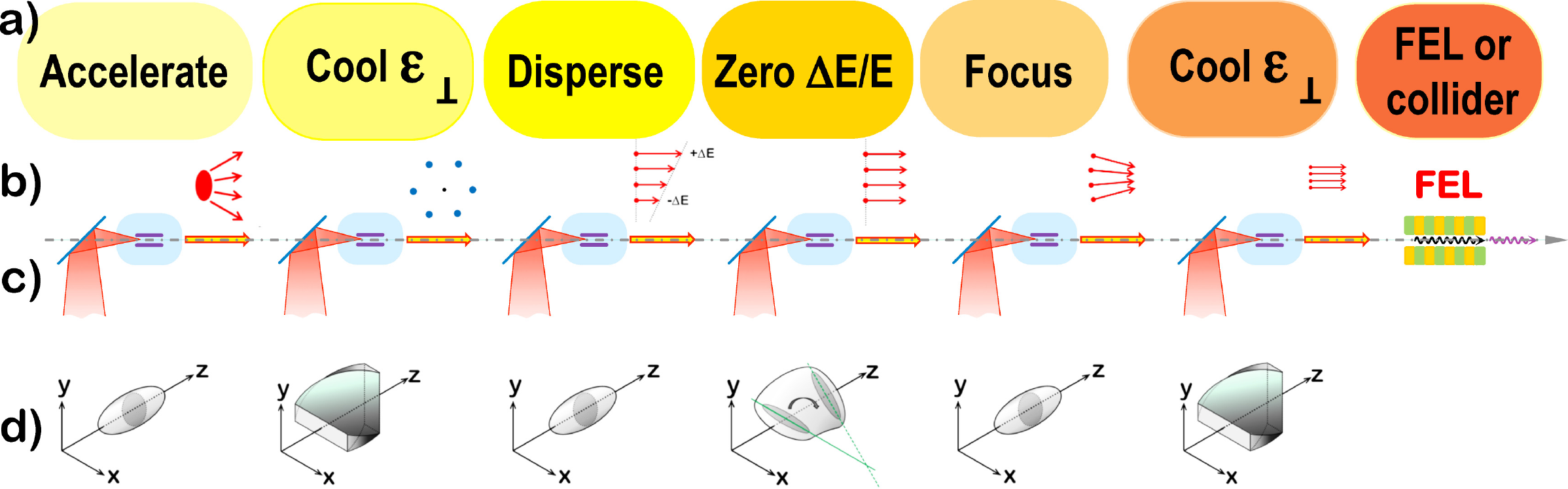}}
  \caption{All sections of the Staged FEL. The row (a) names the individual stages, the row (b) shows the schematics of the beam, the row (c) outlines the geometry of the sections, and the row (d) shows the shape of the laser pulse used for each section.}
\label{overall-fel-collider}
\end{figure}

The entire layout of the Staged FEL, with all sections, shapes of the beams, and shapes of the laser pulse, is shown in Fig.~\ref{overall-fel-collider}. 

As we mentioned above, this is a concept that is enabled by fast synchrotron radiation cooling of the transverse emittance of the beam in the solenoidal field created in the plasma bubble by the screw-shaped laser pulse investigated in \cite{our-paper}. It also relies on a monochromatization stage, which involves a hypothesis that needs to be further studied. 

Further development of this concept, verification of feasibility, optimization of parameters, and experimental tests (with laser or e-driver beams) are of course necessary, and we will briefly discuss this in the next section. 

\section{DISCUSSION}

We presented here the concept of a Staged FEL based on plasma acceleration, where we have included dedicated stages for emittance cooling and monochromatization of the accelerated beam. This design approach will allow one to optimize the acceleration and cooling stages separately, achieving maximum performance of the entire system. The cooling is based on the use of a screw-shaped laser pulse that creates a large solenoidal magnetic field in the plasma bubble, and therefore creates fast synchrotron radiation cooling of the transverse beam emittance. The monochromatization stage is based on a similar laser pulse, but with additional asymmetry in it. The presented concept can be also applied to colliders. 

Further design studies are needed in order to develop the concept further. One of the next design steps would involve detailed simulations of the monochromatization stage, deriving a full parameter set of the system, and defining the overall detailed layout of the Staged FEL. Detailed simulations of the cooling stage are also needed, possibly followed by  experimental tests of the cooling, either with a screw-shaped laser beam or with an equivalent electron beam. The tolerances of all kinds, in particular, the difficulties in the control, synchronization, and alignment of all the stages and all the laser pulses, as well as possible spatial overlap of the latter, would need to be studied. We expect that studies of various aspects of this concept will be presented to the community in the near future.

%


\nocite{*}
\bibliographystyle{aipnum-cp}%
\bibliography{sample}%

\end{document}